\documentclass[12pt]{article}
\def\be{\begin{equation}}
\def\ee{\end{equation}}
\def\ber{\begin{eqnarray}}
\def\eer{\end{eqnarray}}
\usepackage{amsfonts,amssymb}
\begin{document}
%\large
\vspace*{1cm}
\begin{center}
{\Large \bf Mein Ruf to Search New Strange $a_2^s(1520)$-meson}\\

\vspace{4mm}

{\large A.A. Arkhipov\\
{\it State Research Center ``Institute for High Energy Physics" \\
 142280 Protvino, Moscow Region, Russia}}\\
\end{center}

\vspace{4mm}
\begin{abstract}
{In this note we propose to search new strange $a_2^s(1520)$-meson
which is a strange partner of the $a_2(1311)$-meson observed in a
three-pion system.}
\end{abstract}

\section*{}

In our previous papers \cite{1,2,3,4,5,6} we have presented the
arguments in favour of that the Kaluza-Klein picture of the world is
confirmed by the experimentally observed hadronic spectra.

In papers \cite{1,2} the nucleon-nucleon dynamics at very low
energies has been studied and we have found that geniusly simple
formula for KK excitations provided by Kaluza-Klein approach gives an
excellent description for the mass spectrum of two-nucleon system. In
articles \cite{3,4} we have presented additional arguments in favour
of Kaluza and Klein picture of the world. In fact, we have shown that
simple formula provided by Kaluza-Klein approach with the fundamental
scale calculated early \cite{1} gives an excellent description for
the mass spectrum of two-pion and three-pion systems. Taking this
line, we have performed an analysis of experimental data on mass
spectrum of the resonance states containing strange mesons and
compared them with the calculated values provided by Kaluza-Klein
scenario \cite{5}. By this way we have found out quite an interesting
correspondence shown below

\centerline{7-storey}
\[
(?)\sigma(650)\in M_7^{\pi\pi}(640-644)\quad\longleftrightarrow \quad
K^*(892)\in M_7^{K\pi}(893-898)
\]

\centerline{15-storey}
\[
f_2(1275)\in M_{15}^{\pi\pi}(1273-1275)\quad\longleftrightarrow \quad
K_2^*(1430)\in M_{15}^{K\pi}(1431-1434)
\]

\centerline{17-storey}
\[
f_{0,2}(1430)\in M_{17}^{\pi\pi}(1435-1438)\quad\longleftrightarrow
\quad K_2^*(1580)\in M_{17}^{K\pi}(1579-1582)
\]

\centerline{18-storey}
\[
f_{0}(1522)\in M_{18}^{\pi\pi}(1518-1520)\quad\longleftrightarrow
\quad K^*(1680)\in M_{18}^{K\pi}(1654-1657)
\]
\newpage
\centerline{19-storey}
\[
f_{0,?}(1580)\in M_{19}^{\pi\pi}(1599-1601)\quad\longleftrightarrow
\quad K_3^*(1780)\in M_{19}^{K\pi}(1730-1733)
\]

\centerline{22-storey}
\[
\eta_{?,2}(1840)\in
M_{22}^{\pi\pi}(1845-1846)\quad\longleftrightarrow \quad
K_0^*(1950)\in M_{22}^{K\pi}(1960-1963)
\]

\centerline{23-storey}
\[
f_{4}(1935)\in M_{23}^{\pi\pi}(1927-1928)\quad\longleftrightarrow
\quad K_4^*(2045)\in M_{23}^{K\pi}(2038-2040)
\]

\centerline{27-storey}
\[
\rho_{5}(2250)\in M_{27}^{\pi\pi}(2256-2257)\quad\longleftrightarrow
\quad K_5^*(2380)\in M_{27}^{K\pi}(2352-2354)
\]
From this correspondence it follows that $K\pi$-system looks like a
system built from two-pion system by replacement of some one pion
with a kaon. In fact, all experimentally observed hadronic states in
$K\pi$-system have the corresponding partners in two-pion system.
However, some hadronic states in two-pion system do not have the
corresponding strange partners in $K\pi$-system experimentally
observed so far. That is why the further study of $K\pi$-system is
quite a promising subject of the investigations.

Concerning three-pion system we have found out that
\begin{equation}
a_2(1311) \in M_{10}^{3\pi}(1309-1313),\qquad a_2(1311) \in
M_{10}^{\rho\pi}(1310-1312).
\end{equation}
Moreover, we predict the strange partner of $a_2$-meson which we
would like to call as $a_2^s$-meson; see Table 17 and Table 21 in
\cite{5}
\begin{equation}
\fbox{$\displaystyle a_2^s(1520) \in M_{10}^{K2\pi}(1517-1523),\qquad
a_2^s(1520) \in M_{10}^{K\rho}(1519-1522)$}\,.
\end{equation}
Apart of isospin $a_2^s(1520)$-meson may have the same quantum
numbers as $a_2(1311)$-meson. We call up to search
$a_2^s(1520)$-meson and other strange partners of the three-pion
states experimentally observed till now. In this respect it seems the
factory with an intensive kaon beams would be a very good device to
realize such programm.

\end{document}